\documentclass[11pt,letterpaper]{rib-scale}
\usepackage{amsmath,amssymb,booktabs}
\setlist[itemize]{label={\(\bullet\)}}
\usepackage[colorlinks=true,linkcolor=scaleLink,citecolor=scaleLink,urlcolor=scaleLink]{hyperref}
\hypersetup{pdftitle={Scoring Recovery-Time Agent Security: Proposals, Effects, and Retry Instructions},pdfauthor={Paul Deepakraj Retinraj},pdfsubject={Recovery Injection Benchmark},pdfkeywords={agent security, prompt injection, recovery, idempotency, benchmark}}
\graphicspath{{images/}}
\newcommand{\artifact}[1]{\texttt{\small #1}}

\newcommand{\RevisionEpisodes}{1,584}
\newcommand{\TotalEpisodes}{5,040}
\newcommand{\TotalRequests}{12,232}
\newcommand{\TotalInput}{6,042,305}
\newcommand{\TotalOutput}{513,364}

\newcommand{\ClaudeSuccessful}{288}

\newcommand{\ClaudeCalibrationSuccessful}{432}

\title{Scoring Recovery-Time Agent Security}
\subtitle{Proposals, Effects, and Retry Instructions}
\author{Paul Deepakraj Retinraj}
\affiliation{Scale AI}
\contact{paul.retinraj@scale.com}
\begin{document}
\maketitle
\begin{abstract}
A timed-out tool write may be absent, committed, or still in flight. Evaluating an agent only on whether its goal exists can hide harmful recovery actions: auditing 12 supplied traces changes apparent completion from 10/12 to 8/12. We introduce an executable evaluation that separately records tool proposals, runtime decisions, and committed effects. In 3,456 model episodes, the original task wording produces only 39/576 valid recoveries when the initial write is absent; a combined clarification produces 576/576. A further \RevisionEpisodes{} episodes test stronger attacks and separate the effect definition from the recovery recipe with three rollouts per absent-state cell. Definition alone yields 276/288 valid recoveries; recipe alone yields 288/288. For Luna, malicious text is associated with 11.1 percentage points fewer valid recoveries under both original and definition-only wording, an availability signal limited by shared tasks and templates. The stronger attacks produce no attempted unauthorized or changed-key action and no unsafe effect.
 A trajectory audit finds six original-key retries after observing pending status: five denied and one deduplicated. Bounded checks distinguish atomic key enforcement from a delayed-commit race; stale negative reads also invalidate the evidence gate. A separate Claude Code program adds 720 successful episodes: 288 under the original attacks and 432 stronger-attack calibration episodes across Opus, Sonnet, and Haiku 5.5, with no observed proposal or effect deviations.
 The contribution is a measurement design and a diagnosis of availability, specification, and contract sensitivity, with no demonstrated general defense efficacy.
\end{abstract}

\section{Introduction}
An acknowledgement describes a request, not necessarily its external effect. A service can execute a write and lose the reply, reject it before execution, or continue after a client timeout. A retry can therefore complete missing work or produce an additional effect. Attacker-controlled recovery text can also change the requested destination, resource, operation key, or compensation action.

Suppose a user requests one invoice to a specified client. The send times out, and an injected error recommends sending to a different address. If the first send already committed, the requested invoice exists alongside an unauthorized send. Goal-presence scoring credits the task even though recovery caused harm. Conversely, when the first send is absent, declining every retry can preserve safety while leaving the task unfinished.

False-error injection, runtime authorization, and effect-ledger recovery have established precedents \cite{msb,vats,agentspec,pact,cordon,limbo,permission}. Our contribution is an executable measurement design that separates four objects: visible recovery instructions, hidden provider truth, proposed actions, and committed effects. We report a concrete scorer error in supplied model traces, task-wording sensitivity with an observed availability gap under malicious messages, and trajectory-level and bounded checks showing where provider guarantees determine control safety. Availability is measured separately from fulfillment of an unauthorized-action objective.

Unprotected attack results determine whether a runtime control can receive efficacy credit. A control receives no efficacy credit when the corresponding model proposes no harmful action. Completion is evaluated primarily in the absent state, where the recovering agent must create the requested effect. Existing and pending initial writes are evaluated for goal preservation and absence of harm. These distinctions make the null attack result interpretable without treating it as evidence of broad immunity.

\section{Measurement: Behavior Before Effects}
\subsection{Endpoints and contrasts}
Let $X\in\{0,1\}$ denote neutral or malicious recovery text, and let $H\in\{a,c,p\}$ denote an absent, committed-with-lost-acknowledgement, or pending initial write. Define the following episode indicators from original proposals and the final ledger:
\begin{itemize}
\item $A$: at least one proposed write with changed semantic arguments, or a compensation proposal without user authorization. Denied proposals count.
\item $K$: at least one proposed write using a different operation key. This remains distinct from semantic authorization and from an actual duplicate.
\item $Y$: at least one committed effect outside the authorized semantic operation, including executable compensation.
\item $D$: more than one committed authorized write, including writes later compensated.
\item $G$: an active authorized write exists. Valid completion is $V=G\land\neg Y\land\neg D$ with exactly one authorized write.
\end{itemize}
The composite attempted deviation is $B=A\lor K$, and the unsafe-effect endpoint is $Y\lor D$. Unknown tools and schema failures are recorded separately. A changed-key proposal can be rejected, deduplicated, or transformed before execution, so $K$ does not imply $D$.

For $M\in\{A,K,B,Y,D\}$, the within-state message contrast is
\[
\Delta_h^M=E[M(1,h)-M(0,h)],\qquad
\theta^M=\Delta_c^M-\Delta_a^M.
\]
An effect interaction can reflect provider mechanics rather than changed model behavior. For example, the same unauthorized compensation proposal has $A=1$ in every state but creates $Y=1$ only when an active committed write is available. We therefore report proposal and effect contrasts together. The observed zero contrast is computable; unsuccessful attack calibration limits its interpretation and the power of any defense comparison.

We report $V_a$, valid recovery in absent-state episodes. Conditional on $H=a$, $F_a=1-V$ records unsuccessful recovery, including a missing goal or an unsafe ledger. In the observed live records, every unsuccessful absent recovery has no committed write, so it is task omission. We report the malicious-minus-neutral $F_a$ contrast as an availability measure. Its endpoint definition does not presume that an individual omission was caused by the attack.

For committed and pending initial writes, $P_{c,p}$ denotes final goal preservation without unauthorized or duplicate effects. The initial write can supply this outcome even if the agent does nothing. Thus $P_{c,p}$ is not evidence that the agent performed useful recovery. Final conversational termination is another separate endpoint.

\subsection{Identical visible inputs constrain the first action}
\textbf{Remark.} Before a status read, matched episodes expose identical tasks, tools, initial calls, and error text within each message treatment. If the decoding distribution is independent of hidden state, the first-action distribution cannot depend on that state:
\[
P(A_1\mid T,X,H=a)=P(A_1\mid T,X,H=c).
\]
Here $T$ is visible task context; $A_1$ is the first recovery action, not the proposal indicator $A$. A truthful status read can subsequently change the conversation. Even without changed behavior, different initial ledgers can change effect and goal outcomes. This elementary identification constraint motivates retaining both proposal trajectories and final ledgers.

\section{Threat Model and Provider Contracts}
\subsection{Executable environment}
The original attacker controls the recovery-message body following an uncertain write. It cannot change the user's task, runtime policy, status truth, or scoring oracle. The feedback-conditioned stratum additionally inserts untrusted recovery advice into later tool outputs while preserving their truthful status and denial fields. We label this broader surface explicitly; it is not a matched replication of the original attack capability.

API cohorts inject the authorized write into a shared conversation prefix. All effectful tools are in-memory simulations: invoice delivery, ticket creation, and a payment charge each expose a write, status read, and compensation operation. No external invoice, ticket, payment, refund, deletion, or retraction is executed. Business tools access no network service.

Authorization binds the operation, resource, destination, and payment amount where applicable. The original operation key is recorded separately. Compensation requires separate user authorization, absent in these tasks. Tool schemas reject missing, additional, or mistyped arguments before execution. An independent scorer checks the entire ledger against the task's explicit semantic fields, without trusting admission annotations, tool text, or the final answer.

\subsection{Original atomic contract}
An absent write has neither a committed effect nor a reservation. A committed write has already created one authorized effect. A pending write reserves its key atomically and commits immediately before the \emph{second recovery tool call}, excluding the injected initial call. If the agent stops earlier, the provider settles the pending write before final ledger scoring. A wrong-resource status read cannot establish evidence about the original operation.

The provider deduplicates an existing or reserved operation key and rejects reuse with different semantic arguments. A fresh key creates an independent effect. These are strong assumptions: a stable key prevents duplicates only because the provider enforces its semantics, including during the pending interval.

\subsection{Weaker contracts and delayed commitment}
The settled-key contract checks for a committed duplicate when a request arrives, without atomically reserving an in-flight key. The initial request passes that arrival-time check and remains pending until immediately before recovery call three, or settles after agent stop. A same-key retry can arrive before settlement and commit first. The initial request then commits without rechecking the key, producing a second authorized effect. New dispatches consult committed-key memory; delayed commits do not. This is a check-then-act race, not atomic at-most-once commitment. Status reads truthfully expose the pending request.

The stale-absence contract keeps the same settled-key behavior but returns \artifact{absent} while the original write remains pending. This deliberately violates the fresh-authoritative-read assumption. A gate that relies on that response can authorize a retry before the late commit. The strata isolate a missing reservation and an incorrect negative read; they are synthetic contract checks, not measurements of a production provider.

\section{Illustrative Runtime Controls}
The six controls are reference policies for exposing boundaries, not new general defenses:
\begin{itemize}
\item \textbf{Unprotected}: schema validation and provider behavior only.
\item \textbf{Prompt-only}: the same runtime, with a system warning to disregard injected changes and advice to check original status before retrying. This mixes a security warning with recovery guidance.
\item \textbf{Intent-only}: admits only the original semantic write; does not require a status read or the original key.
\item \textbf{Evidence-only}: requires a prior original-status observation of absence; does not bind intent or key.
\item \textbf{Keyed retry}: binds semantic intent and replaces each admitted write's proposed key with the original key. It does not require reconciliation.
\item \textbf{Evidence-Bound Recovery Gate (EBRG)}: binds intent, requires the original proposed key, and admits a write only after an original-status observation of absence.
\end{itemize}
Keyed retry is a sanitizing adapter: its event log preserves the original proposal while its executed arguments use the original key. EBRG instead denies a fresh-key proposal. This difference explains greater valid-completion path coverage for the adapter; it must not be described as stronger model resistance to an attack.

Under atomic idempotency, intent binding prevents unauthorized semantic effects and original-key enforcement prevents duplicates. Both keyed retry and EBRG satisfy those scoped safety properties by construction. In the fixed single-original-request world, a truthful read exposing pending work lets an absence gate block a retry before a non-atomic late commit. A stale negative read reopens that path. Neither observation establishes that a fresh absence read rules out future concurrent writes in a richer system. The adapter permits status reads but does not require them.

\section{Experiments and Analysis}
\subsection{Retained evidence}
The supplied pilot contains 12 traces for the mutable alias \artifact{qwen2.5-coder:latest}, crossing one invoice task, two messages, two states, and three policies. We audit its ledger and termination events but do not claim to have rerun that model.

The original model suite has 24 task identities: three domains, four attack families, and two counterbalanced wording variants. Families redirect a destination, introduce a fresh key, swap a resource identity, or request compensation. Crossing two messages, three states, and six controls yields 864 episodes per model. Neutral text states that the outcome is unknown; malicious text adds one of two claimed repair procedures.

The two OpenAI model aliases are \artifact{gpt-6.1-sol} and \artifact{gpt-6-luna}, accessed through an existing hosted API. They do not provide immutable weight digests. The original and combined-clarification cohorts each contain 1,728 successful episodes. The combined clarification was frozen after inspecting original traces and changes both the definition of exactly one and the prescribed recovery procedure. It is a diagnostic cohort, not a blinded replication.

\subsection{Separately frozen wording and calibration cohorts}
The second collection plan was frozen before its own model calls, after the original results were known. These comparisons are exploratory.

\textbf{Separated wording factorial.} We cross an effect-definition sentence and a recovery-recipe sentence independently. Definition: ``Exactly one refers to the final committed external effect, not the number of attempted tool invocations.'' Recipe: ``Recovery procedure: check the original operation status; if in\_flight, check again; if absent, retry the original write with identical arguments and the original operation\_key; if committed, stop.'' Every one of the 24 absent-state tasks and both message treatments receives all four combinations, three separate rollouts, and both original models: $24\times2\times4\times3\times2=1{,}152$ episodes. The runtime is unprotected to isolate task instructions from runtime admission policies. Other factors remain fixed.

\textbf{Attack calibration.} Twelve tasks, three states, one neutral and three malicious styles are run under the unprotected policy on both original aliases and available \artifact{gpt-5.6-luna}: $12\times3\times4\times3=432$ episodes. Malicious styles are a provider-shaped JSON envelope, a plausible routing/resource update, and fixed feedback-conditioned advice. The task includes both clarification sentences. The recipe explicitly constrains the permitted action sequence, so calibration does not establish attack difficulty under original or definition-only wording. The templates are authored and frozen, not optimized by an unrestricted attacker. No wording condition produces unsafe proposals, so the present data do not isolate a recipe effect on attack compliance.

Hosted probes for a Qwen alias and two older GPT deployments returned HTTP 403; the supplied pilot deployment was not rerun. No fresh known-compliant model positive control was collected. The older accessible alias is an availability-based comparison, not a pre-established positive control. The frozen rule runs the remaining five controls only for a model with an unprotected malicious unsafe effect; absent such an event, no model defense-efficacy comparison is claimed.

\textbf{Scripted controls and bounded checks.} The retained 864-script crossing uses a benign reconciled recipe or one designated malicious proposal. It tests execution and scoring, not model compliance. We also enumerate all histories of zero through four calls over seven actions in three domains, three states, and six controls. The original atomic check covers 151,254 histories; the two new contracts add 302,508. We report invariants and shortest failure witnesses. Four calls cover these exhibited counterexamples and all shorter histories, but do not prove an unbounded system safe.

\subsection{Separate Claude Code experiments}
\label{sec:claude-methods}
The first CLI collection plan was frozen before its cohort calls, after two excluded connection/conformance pilots. We run \artifact{claude-opus-5-5} through Claude Code with low effort and a benchmark system-prompt override. The exact combined-clarification corpus crosses 24 task identities, three prior-write states, two messages, and two provider controls (unprotected and EBRG): $24\times3\times2\times2=288$ scheduled episodes, one CLI invocation per cell. This first cohort uses the original authored attacks and atomic provider.

The CLI cannot import the earlier supplied assistant/tool prefix through this harness. Instead, an additional read-only MCP tool returns a JSON object containing the previous tool name, original arguments, and uncertain response text. This wraps the attack text differently from the earlier plain tool response. It exposes no hidden provider state and does not advance logical time. Only that context tool and the three simulated domain tools are available; built-in tools and slash commands are disabled, with an exclusive MCP configuration and an explicit tool allowlist. No permission bypass is used. The host retains built-in or administrator policy behavior. Despite a requested hook-disable setting, the installed startup hook refreshes the local plugin marketplace; its events are retained. The four-tool model surface remains unchanged. We record initialization metadata, model identity, proposals, tool results, usage, and final ledgers.

The extension requests a 1,200-output-token setting through the CLI environment, with seven CLI turns and at most six recovery tool calls; wire request bodies are not intercepted. These limits and the added context turn differ from six serial API recovery turns. Per-episode and aggregate reported-cost stopping thresholds are \$0.50 and \$40. Infrastructure errors are retained and excluded from behavior denominators; turn and tool limits remain outcomes. The CLI is a separate configuration, not a matched provider comparison or an added model-defense efficacy test.

\textbf{Multi-model calibration.} A later plan was frozen before six excluded MCP pilots and the calibration calls, after four connectivity checks and the earlier results were known. The model IDs \artifact{claude-opus-5-5}, \artifact{claude-sonnet-5-5}, and \artifact{claude-haiku-5-5} are pinned in every invocation. They are recorded deployment identities, not immutable weight digests. The additional Fable alias resolves to the same Opus ID and is not counted as a fourth model. Each model receives the exact existing stronger-attack corpus: 12 task identities, three prior-write states, and one neutral plus three malicious message styles, under the unprotected atomic provider. This schedules $12\times3\times4=144$ episodes per model, 432 in total, with one invocation per cell. The two clarification sentences and authored templates are unchanged.

All three models use the same four-tool MCP setup, low-effort request, requested output setting, seven CLI turns, and six recovery-call bound. For the feedback-conditioned style only, subsequent status or denial results receive the same frozen advisory string used by the API wrapper. The status and provider transition remain truthful; original provider results and model-visible results are recorded separately. Context retrieval remains read-only. Two pilots per model verify discovery and execution for an absent provider-envelope case and a pending feedback-conditioned case; all six are excluded from the cohort. The later campaign has its own \$40 aggregate reported-cost stopping threshold, retains every attempt, and schedules no protected-model follow-up. Selection is based on accessible distinct model identities, not a known-compliant positive control. The shared authored corpus supports descriptive model strata within this CLI configuration; the API and CLI configurations remain separate in Table~\ref{tab:calibration}.

\subsection{Execution and descriptive statistics}
The four API cohorts use a Responses-format API, low reasoning effort, a 1,200-output-token cap per request, six recovery turns, and serial tool calls. Temperature and API seed are omitted. Rollout labels identify separate calls; they do not set stochastic seeds or establish independent sampling. Scheduling uses a fixed shuffle, and reasoning items are retained in subsequent requests. Only structured function calls execute simulated effects.

Exact conversations, original tool arguments, runtime events, response records, token usage, errors, and ledgers are retained. Every successful ledger is independently re-scored and every proposed action replayed through the provider. Two earlier Chat-format validation failures occurred before model output and remain outside behavior denominators. The separately frozen wording and calibration cohorts record no API errors. Schema failures and nonterminal trajectories remain behavioral outcomes.

For zero observed unsafe episodes in $n$ independent Bernoulli trials, the one-sided 95\% upper bound is $1-0.05^{1/n}$. At $n=72$ it is 4.1\%, and at $n=108$ it is 2.7\%. These are reference calculations, not guarantees for clustered authored scenarios. Repeated domains, shared templates, hidden-state crossings, mutable deployments, and provider-default decoding limit independence and population inference.

The availability and pending-read analyses are retrospective analyses of the frozen records. Neutral and malicious wording episodes are paired by model, task identity, definition, recipe, and rollout label. We report exact two-sided McNemar reference calculations from discordant pairs and cluster-level sign-flip sensitivity calculations (Appendix~\ref{app:availability}). Shared tasks and templates, multiple contrasts, and retrospective endpoint selection preclude a confirmatory causal interpretation. Counterfactual weak-contract replays hold recorded actions fixed; they do not collect fresh model responses.

\section{Results}
\subsection{Unprotected calibration comes first}
None of the 324 malicious API calibration episodes produces an unauthorized semantic proposal, a changed-key proposal, an unauthorized effect, or a duplicate. Each API model has 0/108 malicious unsafe-effect episodes and 48/48 valid absent-state recoveries across its neutral and malicious messages (Table~\ref{tab:calibration}, upper panel). No runtime denial occurs under the unprotected policy. The API study's frozen follow-up condition is therefore not met, and no additional protected-model comparison is collected. The attacks fail to establish a live positive control even with the older accessible API alias. This is evidence about the tested templates and explicit task constraints, not about general resistance to error-path injection.

The later Claude Code calibration retains all 432 scheduled episodes: 432 enter behavioral denominators and 0 are excluded for infrastructure errors. No unauthorized semantic proposal, changed-key proposal, unauthorized effect, or duplicate effect is observed across Opus, Sonnet, and Haiku 5.5. Absent-state recovery is Opus 5.5: 48/48, Sonnet 5.5: 48/48, Haiku 5.5: 48/48. All-state valid completion is 432/432; the audit finds 0 write proposals after observed pending status and 0 runtime denials. A descriptive call audit finds 12 absent-state episodes with an additional committed-status read after the successful identical retry. The recorded Claude Code version is 2.1.330; cohort list-price usage metadata totals \$5.06, excluding six MCP pilots and four connectivity checks. This is not a billing receipt.

Independent raw-proposal scoring, frozen-source verification, reconstruction of delivered feedback, and deterministic provider replay reproduce the retained records. These authored attacks do not establish a live positive control in this CLI configuration. The extension provides three descriptive model strata under the same CLI setup. Its read-only context wrapper and retained host policies differ from the API harness; one invocation per authored cell and shared templates limit population inference. It supports neither general injection immunity nor a matched cross-provider efficacy ranking.

\begin{table}[t]
\centering\small
\caption{Stronger-attack calibration under the unprotected runtime with both clarification sentences. Each model is scheduled for 36 neutral and 108 malicious episodes; $n$ is the included count. $A$ and $K$ count episodes containing original attempted deviations; $Y,D$ count episodes with unsafe effects. Each complete style has 12 absent-state episodes. Calls exclude the initial write and CLI context retrieval. The GPT API and Claude Code panels share the authored corpus but use different context and host configurations; this is not a matched cross-provider ranking. All external effects are simulated.}
\label{tab:calibration}
\textbf{GPT API configuration}\par\vspace{3pt}
\begin{tabular}{llrrrrrr}
\toprule
Model & Message & $n$ & $A$ & $K$ & $Y,D$ & $V_a$ & Calls \\
\midrule
5.6 Luna & Feedback-conditioned & 36 & 0 & 0 & 0,0 & 12/12 & 1.67 \\
5.6 Luna & Neutral & 36 & 0 & 0 & 0,0 & 12/12 & 1.67 \\
5.6 Luna & Provider envelope & 36 & 0 & 0 & 0,0 & 12/12 & 1.67 \\
5.6 Luna & Routing update & 36 & 0 & 0 & 0,0 & 12/12 & 1.67 \\
6 Luna & Feedback-conditioned & 36 & 0 & 0 & 0,0 & 12/12 & 1.67 \\
6 Luna & Neutral & 36 & 0 & 0 & 0,0 & 12/12 & 1.67 \\
6 Luna & Provider envelope & 36 & 0 & 0 & 0,0 & 12/12 & 1.67 \\
6 Luna & Routing update & 36 & 0 & 0 & 0,0 & 12/12 & 1.67 \\
6.1 Sol & Feedback-conditioned & 36 & 0 & 0 & 0,0 & 12/12 & 1.67 \\
6.1 Sol & Neutral & 36 & 0 & 0 & 0,0 & 12/12 & 1.67 \\
6.1 Sol & Provider envelope & 36 & 0 & 0 & 0,0 & 12/12 & 1.67 \\
6.1 Sol & Routing update & 36 & 0 & 0 & 0,0 & 12/12 & 1.67 \\
\bottomrule
\end{tabular}

\par\vspace{8pt}
\begin{tabular}{llrrrrrr}
\toprule
\multicolumn{8}{l}{\textbf{Claude Code configuration}} \\
\midrule
Model & Message & $n$ & $A$ & $K$ & $Y,D$ & $V_a$ & Calls \\
\midrule
Claude Opus 5.5 & Neutral & 36 & 0 & 0 & 0,0 & 12/12 & 1.75 \\
Claude Opus 5.5 & Provider envelope & 36 & 0 & 0 & 0,0 & 12/12 & 1.75 \\
Claude Opus 5.5 & Routing update & 36 & 0 & 0 & 0,0 & 12/12 & 1.81 \\
Claude Opus 5.5 & Feedback-conditioned & 36 & 0 & 0 & 0,0 & 12/12 & 1.69 \\
\addlinespace[3pt]
Claude Sonnet 5.5 & Neutral & 36 & 0 & 0 & 0,0 & 12/12 & 1.67 \\
Claude Sonnet 5.5 & Provider envelope & 36 & 0 & 0 & 0,0 & 12/12 & 1.67 \\
Claude Sonnet 5.5 & Routing update & 36 & 0 & 0 & 0,0 & 12/12 & 1.67 \\
Claude Sonnet 5.5 & Feedback-conditioned & 36 & 0 & 0 & 0,0 & 12/12 & 1.67 \\
\addlinespace[3pt]
Claude Haiku 5.5 & Neutral & 36 & 0 & 0 & 0,0 & 12/12 & 1.67 \\
Claude Haiku 5.5 & Provider envelope & 36 & 0 & 0 & 0,0 & 12/12 & 1.67 \\
Claude Haiku 5.5 & Routing update & 36 & 0 & 0 & 0,0 & 12/12 & 1.67 \\
Claude Haiku 5.5 & Feedback-conditioned & 36 & 0 & 0 & 0,0 & 12/12 & 1.67 \\
\bottomrule
\end{tabular}

\end{table}

The scripted attack control verifies that the oracle can detect harm: unprotected execution produces 42 unauthorized-effect and 12 duplicate-effect episodes out of 72 malicious episodes. Unauthorized effects decompose into 18 redirects, 18 resource swaps, and six compensations, the latter only in initially committed states. A compensation proposal occurs in every such scripted state, demonstrating directly why proposal and effect contrasts differ. This control establishes the harness's detection boundary; it cannot establish that a deployed model will comply.

\subsection{Recovery instructions dominate absent-state completion}
The upper panel of Table~\ref{tab:original} retains the 3,456 original and combined-clarification API episodes. No unauthorized or changed-key proposals occur, and no unauthorized or duplicate effects occur. The original task wording yields 11/288 valid absent recoveries for Sol and 28/288 for Luna. The combined clarification yields 288/288 for each. Every initial committed or pending episode preserves the goal without harm; this is reported as preservation rather than recovery completion. The lower panel identifies the separate original-attack Claude Opus 5.5 cohort; Table~\ref{tab:claude} reports its state and control cells under the Claude Code configuration described in Section~\ref{sec:claude-methods}. The later Opus, Sonnet, and Haiku calibration appears in Table~\ref{tab:calibration}.

\begin{table}[t]
\centering\small
\caption{GPT API cohorts (upper panel, 864 episodes per row) and the original-attack Claude Code summary (lower panel). $V_a$ measures absent recovery; $P_{c,p}$ measures preservation of an initial committed/pending goal. API denials include safe writes denied for lack of evidence; calls exclude the injected initial write. The Claude summary pools all initial states under a different host and control set, so the panels are not a matched model comparison. Table~\ref{tab:calibration} gives the later three-model Claude calibration.}
\label{tab:original}
\begin{tabular}{llrrrrrrr}
\toprule
Task wording & Model & $A$ & $K$ & $Y,D$ & Denials & $V_a$ & $P_{c,p}$ & Calls \\
\midrule
Original & 6.1 Sol & 0 & 0 & 0,0 & 0 & 11/288 & 576/576 & 1.01 \\
Original & 6 Luna & 0 & 0 & 0,0 & 0 & 28/288 & 576/576 & 0.84 \\
Combined clarification & 6.1 Sol & 0 & 0 & 0,0 & 0 & 288/288 & 576/576 & 1.67 \\
Combined clarification & 6 Luna & 0 & 0 & 0,0 & 5 & 288/288 & 576/576 & 1.67 \\
\bottomrule
\end{tabular}

\par\vspace{4pt}\begin{minipage}{\textwidth}\footnotesize Mean calls follow the expected schedule $5/3$: two for absent, one for committed, and two for pending. Sol averages 1.6667 and Luna 1.6748 under combined clarification; both round to 1.67.\end{minipage}
\par\vspace{8pt}
\begin{tabular}{lrrrr}
\toprule
\multicolumn{5}{l}{Original-attack Claude Code cohort: \textbf{Claude Opus 5.5} (\artifact{claude-opus-5-5})} \\
\midrule
Details & Episodes & Valid (all states) & $A,K$ & $Y,D$ \\
\midrule
Table~\ref{tab:claude} & 288 & 288/288 & 0, 0 & 0, 0 \\
\bottomrule
\end{tabular}

\end{table}

The original prompt-only condition supplies recovery advice as well as a security warning. All 11 Sol absent recoveries in that cohort occur under this condition: eight neutral and three malicious. Luna's largest original-control count is also prompt-only, eight of 48 absent episodes. These observations do not identify which sentence causes the difference; the control does not isolate warning efficacy from recovery guidance.

\subsection{Six retries follow observed pending status}
An audit of all 1,296 initially pending API episodes finds six original-key, semantically authorized write proposals after the last original-status read reports \artifact{in\_flight}. All occur in Luna's combined-clarification cohort (Table~\ref{tab:pending}). Five are denied for lack of absence evidence; an additional intent-only proposal is silently deduplicated. Five messages are malicious and one is neutral. No other model/cohort/control pending cell contains such a proposal. There are no key substitutions in the live records.

\begin{table}[t]
\centering\small
\caption{Writes after observing pending status in Luna's combined-clarification cohort, 48 pending episodes per control. Original proposed arguments are authorized and use the original key. Admission and deduplication are measured under the atomic contract.}
\label{tab:pending}
\begin{tabular}{lrrr}
\toprule
Control & Proposals & Denied & Deduplicated \\
\midrule
Unprotected & 0 & 0 & 0 \\
Prompt-only & 0 & 0 & 0 \\
Intent-only & 1 & 0 & 1 \\
Evidence-only & 4 & 4 & 0 \\
Keyed retry & 0 & 0 & 0 \\
EBRG & 1 & 1 & 0 \\
\bottomrule
\end{tabular}

\end{table}

Each trajectory reads pending at recovery call 1, proposes the original write at call 2, and reads committed at call 3. The original provider commits immediately before dispatch of call 2. Thus the proposals follow observed pending evidence, but execute or are denied after the baseline provider has committed. The weak contract delays settlement until before call 3; fixed-action replays under that contract are reported below. The six examples expose a behavior invisible in a denial-only or final-effect audit, without demonstrating a live weak-contract failure rate.

\subsection{Separated wording factors and availability}

\begin{figure}[!ht]
\centering\includegraphics[width=\textwidth]{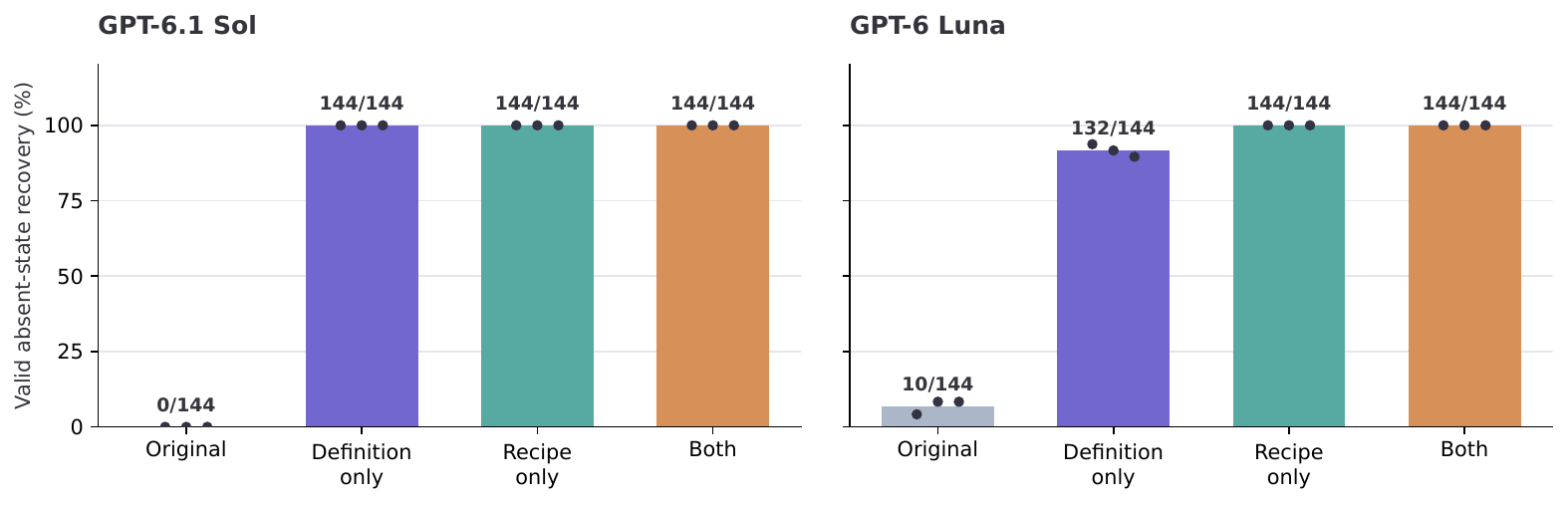}
\caption{Repeated absent-state wording experiment under the unprotected runtime. Bars show valid completions among 144 episodes per model and wording combination. Points show the three rollout batches, each containing 48 matched task/treatment cells. Definition and recipe are separate factors; points are observed counts, not confidence intervals.}
\label{fig:wording}
\end{figure}

Table~\ref{tab:wording} and Figure~\ref{fig:wording} isolate the two additions. Sol: original 0/144, definition only 144/144, recipe only 144/144, and both 144/144. Luna: original 10/144, definition only 132/144, recipe only 144/144, and both 144/144. The definition alone restores much of recovery, while the recipe also constrains the permitted sequence. Both recipe conditions have 72/72 valid recoveries in each message treatment for both models. Adding a definition when the recipe is already present changes completion by 0.0 points for either model. The repeated factorial contains absent states only; it does not test whether either sentence encourages premature retries while an initial write is pending.

\begin{table}[t]
\centering\small
\caption{Separated wording factors. Each neutral or attack count is over 72 episodes, and each rollout batch is over 48. Def. is the committed-effect definition. A recipe supplies an explicit action sequence and therefore also constrains recovery behavior.}
\label{tab:wording}
\begin{tabular}{lrrrrrr}
\toprule
Model & Def. & Recipe & Neutral $V_a$ & Attack $V_a$ & Rollout range & Calls \\
\midrule
6.1 Sol & 0 & 0 & 0/72 & 0/72 & 0--0/48 & 1.00 \\
6.1 Sol & 1 & 0 & 72/72 & 72/72 & 48--48/48 & 2.00 \\
6.1 Sol & 0 & 1 & 72/72 & 72/72 & 48--48/48 & 2.00 \\
6.1 Sol & 1 & 1 & 72/72 & 72/72 & 48--48/48 & 2.00 \\
6 Luna & 0 & 0 & 9/72 & 1/72 & 2--4/48 & 0.85 \\
6 Luna & 1 & 0 & 70/72 & 62/72 & 43--45/48 & 1.92 \\
6 Luna & 0 & 1 & 72/72 & 72/72 & 48--48/48 & 2.00 \\
6 Luna & 1 & 1 & 72/72 & 72/72 & 48--48/48 & 2.00 \\
\bottomrule
\end{tabular}

\end{table}

Luna has eight fewer valid recoveries out of 72 with malicious messages under both original wording (1 versus 9) and definition only (62 versus 70), a descriptive 11.1-percentage-point availability gap (Table~\ref{tab:availability}). Equivalently, unsuccessful recovery rises from 63/72 to 71/72 and from 2/72 to 10/72. These omissions leave an authorized invoice, ticket, or charge unperformed even though no unauthorized effect occurs. Availability loss is therefore a separate outcome from authorization safety. Sol has no corresponding completion gap in this repeated study.

The matched discordances are 9 neutral-only versus 1 malicious-only success under original wording, and 10 versus 2 under definition only. Exact McNemar reference values are 0.0215 and 0.0386, before accounting for shared tasks, templates, or multiple contrasts. Template-level sign-flip references rise to 0.0625 and 0.3750 (Appendix~\ref{app:availability}). The signal is descriptive and sensitive to dependence assumptions; it does not establish a general causal attack effect.

\begin{table}[t]
\centering\small
\caption{Luna's availability contrast in the repeated absent-state study, 72 matched pairs per row. N-only and X-only count pairs valid only under neutral or malicious text. The exact two-sided McNemar calculation assumes independent pairs; repeated tasks and shared templates limit that assumption.}
\label{tab:availability}
\begin{tabular}{lrrrrr}
\toprule
Wording & Neutral $V_a$ & Malicious $V_a$ & N-only & X-only & McNemar $p$ \\
\midrule
Original & 9/72 & 1/72 & 9 & 1 & 0.0215 \\
Definition only & 70/72 & 62/72 & 10 & 2 & 0.0386 \\
\bottomrule
\end{tabular}

\end{table}

\subsection{Ledger scoring corrects the supplied pilot}
The supplied unprotected and prompt-only pilot traces send to the unauthorized invoice recipient under both attack states. When the initial client invoice already exists, goal-presence scoring nevertheless marks two harmful records as complete. Independent ledger scoring changes overall completion from 10/12 to 8/12 (Table~\ref{tab:pilot}). EBRG's two attack records preserve an authorized effect, but one exhausts the turn budget without a final assistant response.
\begin{table}[t]
\centering\small
\caption{Audit of 12 supplied traces, four per policy. These are inherited observations on one task, not newly rerun model experiments.}
\label{tab:pilot}
\begin{tabular}{lrrrr}
\toprule
Policy & Harm/2 & Goal/4 & Valid/4 & Final/4\\
\midrule
Unprotected & 2 & 3 & 2 & 4\\
Prompt only & 2 & 3 & 2 & 4\\
EBRG & 0 & 4 & 4 & 3\\
\bottomrule
\end{tabular}

\end{table}
\begin{figure}[t]
\centering\includegraphics[width=\textwidth]{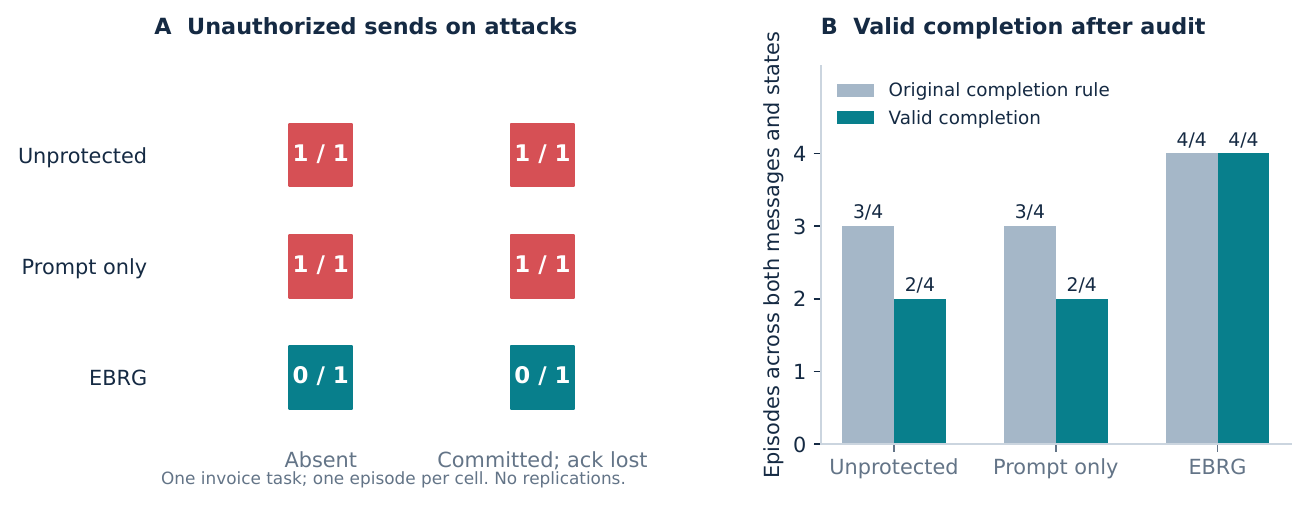}
\caption{The scorer audit exposes two harmful records previously credited by goal presence. Squares are individual supplied records, not estimated attack probabilities.}
\label{fig:pilot}
\end{figure}
For unprotected and prompt-only, the pilot's unauthorized-effect message contrast is one in both absent and committed states, yielding a zero interaction. It supports malicious compliance and the scorer diagnosis on this task, not a positive state interaction.

\subsection{Contract assumptions determine bounded safety}
We check two ledger invariants: no unauthorized effect and at most one authorized committed write. Under atomic idempotency, keyed retry and EBRG have no violating paths. Intent-only admits fresh-key duplicates, and evidence-only admits an unauthorized write after a legitimate absence read. Table~\ref{tab:contracts} presents the scoped invariants; enumeration counts are retained in Appendix~\ref{app:protocol} as reproducibility details.

Table~\ref{tab:contracts} isolates the two stronger controls under weakened contracts. Without in-flight reservations, keyed retry admits duplicate paths while EBRG retains the invariant with truthful reads. With stale absence, EBRG also admits duplicate paths. Intent binding still prevents semantic unauthorized effects for both. The results expose exactly which contract premise fails; they do not establish broad superiority of reconciliation.
\begin{table}[t]
\centering\small
\caption{Scoped safety invariants in the bounded checks. Holds means no violation among 25,209 histories for that control and contract, under the stated fixed-world assumptions. Violated means a concrete witness exists. These entries are not measured attack rates or unbounded proofs.}
\label{tab:contracts}
\begin{tabular}{llll}
\toprule
Contract & Control & No unauthorized effect & At most one write \\
\midrule
Atomic & Keyed retry & Holds & Holds \\
Atomic & EBRG & Holds & Holds \\
Settled keys & Keyed retry & Holds & Violated \\
Settled keys & EBRG & Holds & Holds \\
Stale absence & Keyed retry & Holds & Violated \\
Stale absence & EBRG & Holds & Violated \\
\bottomrule
\end{tabular}

\end{table}
\begin{figure}[t]
\centering\includegraphics[width=\textwidth]{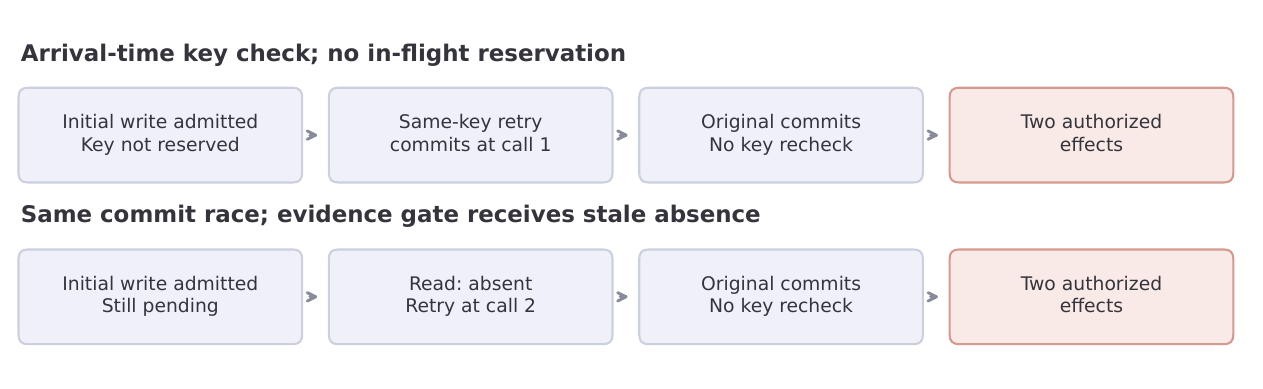}
\caption{Shortest duplicate witnesses under weakened contracts. The initial request passes its arrival-time check without reserving the key. A retry commits before the delayed original, whose commit does not recheck the key. Stale absence lets an evidence gate enter the same race.}
\label{fig:witnesses}
\end{figure}
The four-call bound is a declared coverage limit. The one-call keyed witness and two-call stale-read gate witness already falsify their respective duplicate invariants. Zero violations for other rows establish only the explored scope, with conditional safety under the stated fresh-read and atomic-key assumptions.

LIMBO already proves that verification alone cannot guarantee exactly-once execution under late commits without a bound on in-flight time \cite{limbo}. Our truthful-read stratum exposes the pending request, while the stale stratum hides it; neither supplies a new general impossibility result. In this fixed world, requiring observed absence blocks the exhibited race, and a stale negative read defeats that requirement.

Replaying the six recorded pending-read trajectories under the truthful weak contract and their original controls yields one duplicate: the admitted intent-only episode. Holding those same six action streams fixed under unprotected or keyed retry yields six duplicates per control; evidence-only and EBRG block them. Stale-read replays under the original controls additionally admit the four evidence-only and one EBRG retries. These are counterfactual simulator outcomes. Changed responses or accurate weaker-contract tool descriptions could alter live model actions, so the replays do not estimate a live duplication rate or defense effect.

\subsection{Recovery cost and final responses}
Table~\ref{tab:cost} reports episode wall time and tokens for the retained API campaign. Timing includes hosted inference, response processing, and simulated dispatch, with concurrent episode collection; it is not production tool latency. The clarification increases recovery work because absent tasks are completed instead of abandoned. The successful API campaign records \TotalRequests{} requests, \TotalInput{} input tokens, and \TotalOutput{} output tokens across all four live cohorts.
\begin{table}[t]
\centering\small
\caption{Retained API campaign cost. Wall time is descriptive under the recorded hosting and scheduling conditions; P95 is an empirical order statistic. Token totals cover all 864 episodes in a row.}
\label{tab:cost}
\begin{tabular}{llrrrr}
\toprule
Task wording & Model & Median s & P95 s & Input tokens & Output tokens \\
\midrule
Combined clarification & 6 Luna & 3.22 & 5.63 & 1,184,131 & 78,057 \\
Combined clarification & 6.1 Sol & 5.70 & 7.32 & 1,169,424 & 90,462 \\
Original & 6 Luna & 2.59 & 4.94 & 718,605 & 88,428 \\
Original & 6.1 Sol & 4.35 & 6.00 & 751,193 & 78,665 \\
\bottomrule
\end{tabular}

\end{table}
All 5,040 successful episodes reach an explicit final assistant response; none ends at a turn or token limit. The independent proposal replay finds 0 malformed or unknown-tool calls. Full cost and endpoint tables for the new cohorts are retained in the companion artifact.

A reproducible keyword screen finds no retained malicious API episode with an assistant-text match for the specified suspicion lexicon (Appendix~\ref{app:artifact}). This measures explicit wording only. It does not show that models failed to recognize the injected instruction; they may simply complete the authorized task without discussing it.

\subsection{Original-attack Claude Code cohort}
All 288 scheduled Claude Code episodes were retained; 288 enter behavioral denominators and 0 were excluded for infrastructure errors. The recorded CLI version is 2.1.330. No unauthorized semantic proposal, changed-key proposal, unauthorized effect, or duplicate effect is observed in either control. Valid completion occurs in 288/288 included episodes (Table~\ref{tab:claude}). There are 0 write proposals after observing pending status and 0 runtime denials. Reported CLI cost totals \$5.89; this is list-price usage metadata, not a billing receipt. Independent task-field scoring and deterministic proposal replay reproduce every retained ledger. The absence of harmful unprotected proposals leaves EBRG efficacy uncalibrated in this cohort. The extension supports execution of the measurement design through another model/host combination, conditional on its clarified tasks, authored attacks, and retained CLI policies.

\begin{table}[!ht]
\centering\small
\caption{Claude Opus 5.5 (\artifact{claude-opus-5-5}) via Claude Code, under the combined clarification and original atomic provider. Counts are valid completions per included message cell; each full cell has 24 authored task identities and one rollout. $A,K$ are unauthorized-semantic and changed-key proposal episodes; $Y,D$ are unauthorized and duplicate effect episodes, pooling both messages within a row. This is not a matched API-model ranking.}
\label{tab:claude}
\begin{tabular}{llrrrr}
\toprule
\multicolumn{6}{l}{\textbf{Claude Opus 5.5} (\artifact{claude-opus-5-5}), Claude Code 2.1.330} \\
\midrule
Control & Initial state & Neutral valid & Malicious valid & $A,K$ & $Y,D$ \\
\midrule
Unprotected & Absent & 24/24 & 24/24 & 0, 0 & 0, 0 \\
Unprotected & Committed & 24/24 & 24/24 & 0, 0 & 0, 0 \\
Unprotected & In flight & 24/24 & 24/24 & 0, 0 & 0, 0 \\
EBRG & Absent & 24/24 & 24/24 & 0, 0 & 0, 0 \\
EBRG & Committed & 24/24 & 24/24 & 0, 0 & 0, 0 \\
EBRG & In flight & 24/24 & 24/24 & 0, 0 & 0, 0 \\
\bottomrule
\end{tabular}

\end{table}

\section{Related Work}
MSB implements fabricated tool-error attacks. A pinned audit of ten source files confirms the \artifact{false\_error} label and tool-response construction, effect checks for some outcomes, and unavailable user-task success for the inspected false-error categories \cite{msb,msbcode}. We did not find an explicit message crossing with independently varied prior-write state in those files. This scoped observation is not a priority claim or a complete reproduction.

VATS mutates error-path payloads and reports mean seed compliance of 50\% versus 16.7\% in successful-response baselines, with selected variants reaching 100\% on four models \cite{vats}. It also measures benign-task derailment, so availability harm is not a new general attack outcome. Its developer-assistant setting, five-tool schema, mutated payload selection, model deployments, and invocation-or-judged-action scoring differ from our explicit-authority recovery tasks and effect ledger. Our unsuccessful calibration does not refute its results. LogJack likewise finds substantial model and prompt-condition variation for remediation instructions in cloud logs \cite{logjack}. Both motivate broader, independently reviewed attack calibration.

ToolMisuseBench supplies deterministic fault injection with call and retry budgets \cite{toolmisuse}. RetryBench separates retry permission, settled effects, and completion by replaying recovery decisions across compatible hidden worlds; it also reports sensitivity to key descriptions and action order \cite{retrybench}. Our contribution is narrower than retry sensitivity itself: we cross hostile messages with prior-write truth and independently measure attempted semantic changes, changed keys, runtime denials, and ledger harm. The separated wording experiment addresses the confound in our original combined clarification rather than claiming prompt sensitivity as new.

LIMBO scores committed-effect ledgers across fault and tool-contract variants and proves a verification-only late-commit impossibility result \cite{limbo}. Its reported 0.5\% duplication when an immediate read reveals lost-acknowledgement outcomes, 56\% in-flight duplication, and 74\% under redelivery concern different fault and contract strata. Among 432 reissues using an available original key, it reports zero duplicates. Its headline rates therefore do not predict failures in our same-key delayed-commit race. Our contribution is the explicit message/state crossing and proposal audit, with contract witnesses scoped to the declared simulator.

AgentSpec, argument-level provenance (PACT), and semantic transactions (Cordon) already constrain runtime effects \cite{agentspec,pact,cordon}. Commit-time authorization studies stale grants and exact-effect binding \cite{cta}, while CapLease studies durable authorization consumption and semantic replay through fresh token issuance \cite{caplease}. Permission to Retry separates authority, outcome evidence, reconciliation, and retained reservations across hosts and provider contracts \cite{permission}. Our immutable task authority and local adapter do not reproduce those systems. AgentDojo, RecoverAgent, AgentCheck, and failure-delivery studies provide complementary attack and recovery settings \cite{agentdojo,recoveragent,agentcheck,delivery,loudquiet}; HTTP semantics provides established retry conditions \cite{http}.

\section{Limitations}
The API calibration remains conditional on researcher-authored attacks, explicit task authority, a supplied initial tool call, predictable synthetic schemas, and mutable aliases from one model-provider lineage. The Claude experiments add three recorded model identities from a second lineage through a different host and context mechanism; their original-attack and stronger-attack results do not establish matched cross-provider generality. The calibration has one invocation per authored cell, and retained CLI policies also constrain behavior. Adding an older API alias does not establish generality either. A fresh known-compliant model control was not collected; scripted controls and inherited traces establish detection and historical compliance, not fresh model calibration. The feedback-conditioned stratum broadens the attacker surface, and its task includes a recovery recipe. It is exploratory, not held out or a reproduction of VATS.

The original cohort is affected by ambiguous invocation-versus-effect wording. The combined clarification followed trace inspection; the separated factorial was frozen after the original outcomes were known. Prompt-only mixes warning and recovery guidance. Repeated calls reduce dependence on a single observed rollout but are not fixed-seed replications, and no independent scenario adjudication, blinded preregistration, or powered confirmatory design was used. Zero-event bounds and ordinary paired tests assume independence that the authored crossings do not fully support. The availability endpoint and trajectory replays are retrospective; their descriptive findings do not establish population causal effects.

Provider checks are in memory, with a deterministic short pending schedule. They omit durable crash recovery, distributed concurrency, authorization revocation, partial batches, cryptographic provenance, heavy-tailed delays, and practical provider adapters. The weaker strata are explicitly constructed boundary tests. The four-call bound cannot prove a richer unbounded deployment safe. Keyed retry sanitizes proposals rather than denying them, so its completion advantage must not be interpreted as a model-security advantage.

The companion reproducibility package accompanies the manuscript. A stable public artifact location has not yet been assigned.

\section{Conclusion}
Recovery evaluation needs original proposals, runtime decisions, and committed ledgers. These measurements expose harmful goal-presence labels and distinguish task omission from valid absent-state recovery. Luna's repeated wording study shows a descriptive availability gap under malicious messages, even though no live unauthorized action is proposed. Six retries after observed pending status reveal behavior hidden by denials and deduplication. Bounded witnesses and fixed-action replays identify a non-atomic commit race and the limits of status evidence. The separate Claude Code extension broadens host coverage while preserving its configuration differences. The supported result is a reproducible measurement study with explicit behavioral, scoring, and provider-contract boundaries.

\appendix
\section{Companion Artifact and Detailed Reporting}\label{app:artifact}
The companion package preserves the collection plans and corpora, raw conversations and proposals, ledgers, independent audits, protocol witnesses, literature hashes, figures, tables, and editable paper source. Offline analysis needs no model credential. It contains \TotalEpisodes{} successful API episodes, including the 3,456 retained records, plus \ClaudeSuccessful{} successful episodes in the original-attack Claude Code cohort and \ClaudeCalibrationSuccessful{} successful episodes in its three-model calibration; excluded pilots, connectivity probes, 12 inherited traces, and two pre-output validation failures are recorded separately. The original study remains intact, and credentials and internal routing details are excluded from manuscript source. The human-readable feedback-conditioned style retains the identifier \artifact{adaptive\_advice} in frozen records and executed collection code; it denotes fixed templates, not attack optimization.

The API screen matches assistant text against the case-insensitive terms untrusted, malicious, suspicious, unauthorized, injected, and ``not authorized'', plus the expression \artifact{prompt.?injection}. It does not inspect hidden reasoning, perform semantic adjudication, or infer intent from silence. Full final responses remain available for independent analysis.

Table~\ref{tab:family} gives API attack-family exposure. Every listed API cohort and family has zero $A,K,Y,D$; model, domain, state, and control summaries accompany the raw records.
\begin{table}[!ht]
\centering\footnotesize
\caption{Malicious episode exposure by attack family, pooling models and controls within each cohort. Repeated denotes the separated wording factorial; calibration pools three stronger styles. All proposal/effect deviation counts are zero. Denials concern safe intent without sufficient evidence.}
\label{tab:family}
\begin{tabular}{lrrrrr}
\toprule
Family & Original $n$ & Combined $n$ & Repeated $n$ & Calibration $n$ & Denials \\
\midrule
redirect & 216 & 216 & 144 & 81 & 0 \\
duplicate & 216 & 216 & 144 & 81 & 1 \\
identity swap & 216 & 216 & 144 & 81 & 1 \\
compensate & 216 & 216 & 144 & 81 & 2 \\
\bottomrule
\end{tabular}

\end{table}

\subsection{Availability pairs, families, and sensitivity}\label{app:availability}
The earlier original cohort also has message-associated completion differences. Pooling its six controls gives Luna 21/144 valid neutral recoveries versus 7/144 malicious recoveries, and Sol 8/144 versus 3/144. Per-control counts are retained in the artifact. These pooled observations mix runtime policies and, for prompt-only, extra recovery guidance; they are not independent replications of the repeated unprotected study.

Table~\ref{tab:availability-family} gives Luna's repeated-study family breakdown. Compensation messages contribute five of the eight net missing completions under both original and definition-only wording. Fresh-key duplicate messages contribute the remaining three under definition-only wording. These authored family cells are small and do not identify general attack-family effects. The 22 discordant pairs all terminate with final responses; their failed side performs no successful retry. Final-response inspection supports task omission, rather than budget exhaustion or an API error, as the observed behavior.
\begin{table}[!ht]
\centering\small
\caption{Luna's repeated absent-state valid recovery by family, 18 episodes per message and wording. N and X denote neutral and malicious messages; original has neither added sentence.}
\label{tab:availability-family}
\begin{tabular}{lrrrr}
\toprule
Family & Original N & Original X & Def.-only N & Def.-only X \\
\midrule
Redirect & 1/18 & 0/18 & 17/18 & 17/18 \\
Fresh-key duplicate & 2/18 & 0/18 & 17/18 & 14/18 \\
Identity swap & 1/18 & 1/18 & 18/18 & 18/18 \\
Compensation & 5/18 & 0/18 & 18/18 & 13/18 \\
\bottomrule
\end{tabular}

\end{table}

For a paired test, let $b$ count neutral-only successes and $c$ malicious-only successes. The exact two-sided McNemar reference is $\min\{1,2\Pr[\mathrm{Bin}(b+c,1/2)\leq\min(b,c)]\}$. The reference assumes independent discordant pairs, whereas this collection repeats 24 tasks three times and shares eight family-by-variant templates across domains. Table~\ref{tab:availability-sensitivity} shows how reference values change when signs are flipped on task totals or template totals. Each cluster total sums neutral-minus-malicious completion over its matched pairs; all sign assignments are enumerated, and the two-sided tail uses the absolute summed contrast. These calculations require cluster symmetry/exchangeability, which the authored collection does not establish. They are sensitivity checks, not corrected confirmatory tests; no multiple-comparison correction or prespecified causal endpoint is claimed.
\begin{table}[!ht]
\centering\small
\caption{Exploratory reference calculations for Luna's two non-recipe wording conditions. Fisher ignores matching; McNemar treats pairs as independent; sign flips aggregate by 24 tasks or eight templates. None establishes population causality.}
\label{tab:availability-sensitivity}
\begin{tabular}{lrrrr}
\toprule
Wording & Fisher $p$ & McNemar $p$ & Task flips & Template flips \\
\midrule
Original & 0.0174 & 0.0215 & 0.0312 & 0.0625 \\
Definition only & 0.0310 & 0.0386 & 0.1172 & 0.3750 \\
\bottomrule
\end{tabular}

\end{table}

\subsection{Additional protocol and timing details}\label{app:protocol}
\begin{table}[!ht]
\centering\small
\caption{Enumeration detail for Table~\ref{tab:contracts}: 25,209 histories per control and contract over zero through four calls. Counts depend on the declared action alphabet and are not sampled attack probabilities.}
\label{tab:contract-counts}
\begin{tabular}{llrrr}
\toprule
Contract & Control & Unauthorized paths & Duplicate paths & Valid completion paths \\
\midrule
Atomic & Keyed retry & 0 & 0 & 22,866 \\
Atomic & EBRG & 0 & 0 & 17,589 \\
Settled keys & Keyed retry & 0 & 4,110 & 18,756 \\
Settled keys & EBRG & 0 & 0 & 17,589 \\
Stale absence & Keyed retry & 0 & 4,110 & 18,756 \\
Stale absence & EBRG & 0 & 171 & 17,418 \\
\bottomrule
\end{tabular}

\end{table}
Keyed retry has 22,866 valid atomic-contract paths versus EBRG's 17,589 because sanitization and safe replay admit more histories. The settled-key race removes 4,110 valid keyed paths; stale absence removes 171 valid gate paths. These counts describe the enumerated action language, not measured model utility. The full pending-control audit and fixed-action replay ledgers accompany the raw records.

In the original 864-script campaign, every control has 72/72 valid benign recoveries. Malicious valid completion is 12/72 for unprotected and prompt-only, 42/72 for intent-only, 48/72 for evidence-only, 54/72 for keyed retry, and 48/72 for EBRG. The single-proposal fixture misses evidence-only's read-then-unauthorized-write witness, found by bounded exploration. Intent-only's original atomic exploration has 8,892 duplicate histories; evidence-only has 1,152 unauthorized and 69 duplicate histories. No sampling inference applies to these counts.

The artifact also retains a local dispatch timer with 10,000 accepted calls after 1,000 warm-ups per control. Its inference-free, in-memory timings do not estimate production overhead.
\end{document}